\documentclass[twocolumn,pra,showpacs,nofootinbib]{revtex4}

\usepackage[latin1]{inputenc}
\usepackage{amsmath,amssymb,amscd,bbm,epsfig}
\usepackage[all,cmtip]{xy}

\newcommand{\bignone}{\,}

\newcommand{\tmmathbf}[1]{\ensuremath{\boldsymbol{#1}}}
\newcommand{\tmop}[1]{\ensuremath{\operatorname{#1}}}
\newcommand{\tmtextbf}[1]{{\bfseries{#1}}}
\newcommand{\tmtextit}[1]{{\itshape{#1}}}

\begin{document}

\title{Nakajima-Zwanzig versus time-convolutionless master equation for the
non-Markovian dynamics of a two-level system}
\author{Andrea \surname{Smirne}${}^{a,b}$}
\email{andrea.smirne@unimi.it}
\author{Bassano \surname{Vacchini}${}^{a,b}$}
\email{bassano.vacchini@mi.infn.it}
\affiliation{\mbox {${}^a$Universit{\`a} degli Studi di Milano, Dipartimento di
Fisica, Via Celoria 16, I-20133 Milano, Italy}\\
\mbox{${}^b$INFN, Sezione di Milano, Via Celoria 16, I-20133 Milano, Italy}
}
\date{\today}

\begin{abstract} We consider the exact reduced dynamics of a two-level
system coupled to a bosonic reservoir, further obtaining the exact
time-convolutionless and Nakajima-Zwanzig non-Markovian equations of
motion. The considered system includes the damped and undamped
Jaynes-Cummings model. The result is obtained by exploiting an
expression of quantum maps in terms of matrices and a simple relation
between the time evolution map and time-convolutionless generator as
well as Nakajima-Zwanzig memory kernel. This non-perturbative
treatment shows that each operator contribution in Lindblad form
appearing in the exact time-convolutionless master equation is
multiplied by a different time dependent function. Similarly, in the
Nakajima-Zwanzig master equation each such contribution is convoluted
with a different memory kernel.  It appears that depending on the
state of the environment the operator structures of the two set of
equations of motion can exhibit important differences.
\end{abstract} 
\maketitle

\section{Introduction\label{sec:intro}}

The study of open quantum systems is a wide research area of interest to
various scientific communities, ranging from physicists to chemists and
mathematicians. Its basic theoretical framework is well-understood when a
Markovian description can be applied {\cite{Alicki2007,Breuer2007}}, but
despite important work {\cite{Breuer2007,Weiss2008}} a lot remains to be
clarified and understood when considering non-Markovian dynamics, which is
often mandatory for a realistic approach, also coping with strong coupling
between system and environment. Recently a lot of research work has been
devoted to the subject
{\cite{Budini2004a,Budini2005a,Budini2005b,Budini2006a,Maniscalco2006a,Breuer2006a,Breuer2007a,Vacchini2008a,Ferraro2008a,Kossakowski2008a,Kossakowski2009a,Breuer2008a,Breuer2009a,Piilo2008a,Piilo2009a,Breuer2009c,Vacchini2010b,Chruscinski2010a,Chruscinski2010b,Laine-xxx2,Mazzola-xxx}}.
On the one hand it is not fully clear which is the most general operator
structure of non-Markovian equations of motion, which do provide a
well-defined time evolution and in particular preserves complete positivity.
On the other hand one would like to link, in a possibly intuitive way,
operator structures giving a sensible dynamical evolution with microscopic
information on the physics of the system of interest. For the Markovian case
both these approaches, phenomenological and microscopic, are well understood
and successfully used. The result by Gorini, Kossakowski, Sudarshan and
Lindblad {\cite{Gorini1976a,Lindblad1976a}} provides a robust framework in
order to envisage a possible ansatz for the evolution equations, while the
Markov approximation is often connected with weak coupling, thus making
microscopic approaches more manageable. In considering non-Markovian dynamics,
there is a natural tendency to bring over the intuition gained for the
Markovian case. As it appears, however, the non-Markovian case is much more
subtle and involved, so that in such a way one often runs into inconsistencies
and pitfalls.

In this article we will discuss a realistic physical model, simple enough to
be exactly treated in detail, but already showing typical non-Markovian
features. In particular we not only derive the exact equations of motion,
which provide the counterpart to the Lindblad structure of a Markovian model,
but also obtain a clear-cut connection to the microscopic approach via the two
standard Nakajima-Zwanzig and time-convolutionless techniques. This work
should help in better understanding typical non-Markovian features in simple
but realistic models. We will consider a two-level system coupled first to a
single mode of the radiation field, and later to a bath of harmonic
oscillators, via a Jaynes-Cummings type of interaction. By exploiting the
knowledge of the exact unitary evolution, and therefore of the reduced
dynamics, as well as a suitable matrix representation of the dynamical maps,
we can exhibit the exact time-convolutionless and Nakajima-Zwanzig equations
of motion. It will appear that such equations are given by a sum of Lindblad
terms, each multiplied or convoluted with a different time dependent function.
Moreover the operator structure of these two sets of equations will be shown
to be generally different, also depending on the environmental state. These
exact results, with a clear-cut connection to a microscopic perturbative
derivation, should guide in developing a correct intuition for the approximate
treatment of more involved non-Markovian systems.

The article is organized as follows. In Sect.~\ref{sec:jc} we derive the map
giving the exact reduced dynamics for the Jaynes-Cummings model, further
introducing a simple representation of the map as a matrix with respect to a
convenient basis of operators. In Sect.~\ref{sec:ms} we point out the simple
relations connecting the time evolution map to the time-convolutionless,
time-dependent generator and the Nakajima-Zwanzig memory kernel, respectively.
These relationships are most easily expressed in matrix form. The
corresponding matrices are evaluated for the reduced dynamics of our two-level
system, further coming back to the operator expression of the master equations
and giving the explicit result for a bath in the vacuum state. In
Sect.~\ref{sec:djc} we put forward a similar analysis for the damped two-level
system, making contact with previous results and comparing with the undamped
case.

\section{Jaynes-Cummings model and exact reduced dynamics\label{sec:jc}}

\subsection{Time evolution mapping}

We consider a two-level system coupled to a single mode of the radiation field
according to the total Hamiltonian
\begin{eqnarray}
  H & = & H_S + H_E + H_I,  \label{eq:htot}
\end{eqnarray}
where the system Hamiltonian is given by
\begin{eqnarray}
  H_S & = & \omega_0 \sigma_+ \sigma_-,  \label{eq:hs}
\end{eqnarray}
with $\omega_0$ the transition frequency, $\sigma_+ = |1 \rangle \langle 0|$
and $\sigma_- = |0 \rangle \langle 1|$ the raising and lowering operators of
the two-level system. The Hamiltonian for the single mode of the radiation
field is given by
\begin{eqnarray}
  H_E & = & \omega b^{\dag} b,  \label{eq:he}
\end{eqnarray}
where the creation and annihilation operators $b^{\dag}$ and $b$ obey the
standard bosonic commutation relation. The coupling is in the Jaynes-Cummings
form
\begin{eqnarray}
  H_I & = & g \left( \sigma_+ \otimes b + \sigma_- \otimes b^{\dag} \right), 
  \label{eq:hi}
\end{eqnarray}
so that the considered model can describe e.g. the interaction between a
two-level atom and a mode of the radiation field in electric dipole and
rotating wave approximation. Working in the interaction picture with respect
to the free Hamiltonian $H_S + H_E$,
\begin{eqnarray}
  H_I (t) & = & g \left( \sigma_+ \otimes be^{i \Delta t} +
  \sigma_- \otimes b^{\dag} e^{- i \Delta t} \right),  \label{eq:hit}
\end{eqnarray}
with
\begin{eqnarray}
  \Delta & = & \omega_0 - \omega  \label{eq:delta}
\end{eqnarray}
the detuning between the system and the field mode, it is possible to obtain
the exact dynamics generated by the total Hamiltonian (see e.g. Puri
{\cite{Puri2001}}), and therefore the reduced dynamics of the two-level
system. We express the result exhibiting the unitary evolution operator which
in the basis $\left\{ |1 \rangle, |0 \rangle \right\}$ is given by the
following matrix, whose entries are operators in the Fock space of the
radiation field
\begin{eqnarray}
  U (t) & = & \left(\begin{array}{cc}
    c_{} \left( \hat{n} + 1, t \right) & d \left( \hat{n} + 1, t \right) b\\
    - b^{\dag} d^{\dag} \left( \hat{n} + 1, t \right) & c^{\dag} \left(
    \hat{n}, t \right)
  \end{array}\right),  \label{eq:u}
\end{eqnarray}
where the following operators have been introduced
\begin{eqnarray}
  c \left( \hat{n}, t \right) & = & e^{i \Delta t / 2}  \left[ \cos \left(
  \sqrt{\Delta^2 + 4 g^2 \hat{n}} \frac{t}{2} \right)   \label{eq:cn}
  \right.\\
  &  & \left. - i \Delta \frac{\sin \left( \sqrt{\Delta^2 + 4 g^2 \hat{n}}
  \frac{t}{2} \right) }{\sqrt{\Delta^2 + 4 g^2 \hat{n}}}  \right], \nonumber\\
  d \left( \hat{n}, t \right) & = & - ie^{i \Delta t / 2} 2 g \frac{\sin
  \left( \sqrt{\Delta^2 + 4 g^2 \hat{n}} \frac{t}{2} \right) }{\sqrt{\Delta^2
  + 4 g^2 \hat{n}}},  \label{eq:dn}
\end{eqnarray}
with $\hat{n} = b^{\dag} b$ the number operator. The unitarity of $U \left( t
\right)$ is granted because of the easily verified relation
\begin{eqnarray}
  c^{\dag} \left( \hat{n}, t \right) c \left( \hat{n}, t \right) + \hat{n}
  d^{\dag} \left( \hat{n}, t \right) d \left( \hat{n}, t \right) & = & 1. 
  \label{eq:vincolo}
\end{eqnarray}
Given the unitary evolution of the whole bipartite system one can obtain the
reduced dynamics of the two-level atom simply by taking the partial trace with
respect to the environmental degrees of freedom. If the initial state of
system and environment is factorized, the map giving the reduced dynamics is
completely positive and takes the form
\begin{eqnarray}
  \rho (t)  & = & \tmop{Tr}_E \left[ U (t) \rho
  \left( 0 \right) \otimes \rho_E U^{\dag} (t) \right] 
  \label{eq:defphi}\\
  & \equiv & \Phi (t) \rho \left( 0 \right) . \nonumber
\end{eqnarray}
Taking $U (t)$ as in Eq.~(\ref{eq:u}) and considering an
environmental state commuting with the number operator, $\left[ \rho_E, n
\right] = 0$, so that in particular both the vacuum and a thermal state can be
dealt with, one comes to the following explicit expression for the action of
the map $\Phi (t)$:
\begin{eqnarray}
  \rho_{11} (t) & = & \rho_{00} \left( 0 \right) \left( 1 -
  \alpha (t) \right) + \rho_{11} \left( 0 \right) \beta \left( t
  \right), \nonumber\\
  \rho_{10} (t) & = & \rho_{10} \left( 0 \right) \gamma \left( t
  \right), \nonumber\\
  \rho_{01} (t) & = & \rho_{01} \left( 0 \right) \gamma^{\ast}
  (t), \nonumber\\
  \rho_{00} (t) & = & \rho_{00} \left( 0 \right) \alpha \left( t
  \right) + \rho_{11} \left( 0 \right) \left( 1 - \beta (t)
  \right) .  \label{eq:rd}
\end{eqnarray}
The effect of the interaction with the bath is contained in the time dependent
coefficients $\alpha (t)$, $\beta (t)$ and $\gamma
(t)$, which are given by the following expectation values over
the state of the environment $\rho_E$:
\begin{eqnarray}
  \alpha (t) & = & \langle c^{\dag} \left( \hat{n}, t \right) c
  \left( \hat{n}, t \right) \rangle_E, \nonumber\\
  \beta (t) & = & \langle c^{\dag} \left( \hat{n} + 1, t \right)
  c \left( \hat{n} + 1, t \right) \rangle_E,  \label{eq:abc}\\
  \gamma (t) & = & \langle \mathbbm{} c \left( \hat{n}, t \right)
  c_{} \left( \hat{n} + 1, t \right) \rangle_E . \nonumber
\end{eqnarray}

\subsection{Matrix representation}

Now that we have obtained the completely positive map $\Phi (t)$
giving the exact reduced time evolution of the considered two-level system, we
shall express it for later convenience in matrix form with respect to a
suitable basis of operators in the Hilbert space $\mathbbm{C}^2$, following
{\cite{Andersson2007a}}. Indeed, having fixed a basis $\left\{ X_l \right\}_{l
= 0, 1, 2, 3}$ of operators in $\mathbbm{C}^2$, orthonormal according to
$\tmop{Tr}_S \left[ X^{\dag}_k X_l \right] = \delta_{kl}$, a linear map
$\Lambda$ can be expressed in this basis according to
\begin{eqnarray}
  \Lambda \rho & = & \sum_{k l} L_{k l} \tmop{Tr}_S \left[ X_l^{\dag} \rho
  \right] X_k,  \label{eq:lambda}
\end{eqnarray}
where the matrix of coefficients $L_{k l}$ uniquely associated to the map is
given by
\begin{eqnarray}
  L_{k l}  & = & \tmop{Tr}_S \left[ X^{\dag}_k \Lambda \left( X_l \right)
  \right] .  \label{eq:elle}
\end{eqnarray}
Here and in the following we will use Greek or calligraphic letters to denote
maps, and Roman letters to indicate the corresponding matrix. The convenient
choice of basis for our calculations, in order to later recast the relevant
maps in operator form, is given by $\left\{ X_l \right\}_{l = 0, i} \equiv
\left\{ \frac{1}{\sqrt{2}} \mathbbm{1}, \frac{1}{\sqrt{2}} \sigma_i \right\}$,
where $\sigma_i$ denote the usual Pauli operators. This choice leads to the
following expression for the matrix $F_{kl} (t)$ associated to
the time evolution map $\Phi (t)$:
\begin{equation}
  F (t) =  \left(\begin{array}{cccc}
    1 & 0 & 0 & 0\\
    0 & \gamma_R (t) & \gamma_I (t) & 0\\
    0 & - \gamma_I (t) & \gamma_R (t) & 0\\
    \beta (t) - \alpha (t) & 0 & 0 & \beta \left( t
    \right) + \alpha (t) - 1
  \end{array}\right),  \label{eq:F}
\end{equation}
where the coefficients defined in Eq.~(\ref{eq:abc}) appear, and we denote
with $R$ and $I$ real and imaginary part of a given function:
\begin{eqnarray}
  \gamma & = & \gamma_R + i \gamma_I . \nonumber
\end{eqnarray}
This strategy of associating matrices to maps, already pursued in
{\cite{Andersson2007a}}, and which transforms composition in matrix
multiplication, will turn out to be very convenient to obtain the expressions
of exact equations of motion leading to the dynamics described by
Eq.~(\ref{eq:rd}).

\section{Exact Nakajima-Zwanzig and time-convolutionless master
equations\label{sec:ms}}

\subsection{General expression in terms of time evolution map}

With the aid of the knowledge of the exact time evolution, and using the
representation of maps in terms of matrices, we will now explicitly obtain two
kinds of exact equations of motion for the reduced system's dynamics.

We first consider a master equation in differential form with a generator
local in time, i.e. the time-convolutionless master equation. Assuming the
existence of such a generator $\mathcal{K}_{\tmop{TCL}} (t)$, it
should obey the equation
\begin{eqnarray}
  \dot{\rho} (t) & = & \mathcal{K}_{\tmop{TCL}} (t)
  \rho (t),  \label{eq:mstcl}
\end{eqnarray}
which due to $\rho (t) = \Phi (t) \rho \left( 0
\right)$ is satisfied upon identifying
\begin{eqnarray}
  \mathcal{K}_{\tmop{TCL}} (t) & = & \dot{\Phi} (t)
  \Phi^{- 1} (t),  \label{eq:mappatcl}
\end{eqnarray}
or in terms of matrices
\begin{eqnarray}
  K_{\tmop{TCL}} (t) & = & \dot{F} (t) F^{- 1}
  (t),  \label{eq:matricetcl}
\end{eqnarray}
holding provided the inverse does exist.

On a similar footing one can consider a master equation in integrodifferential
form, given by a suitable memory kernel, corresponding to the Nakajima-Zwanzig
master equation. In this case the memory kernel $\mathcal{K}_{\tmop{NZ}}
(t)$ should obey the convolution equation
\begin{eqnarray}
  \dot{\rho} (t) & = & \left( \mathcal{K}_{\tmop{NZ}} \star \rho
  \right) (t),  \label{eq:msnz}
\end{eqnarray}
so that in view of Eq.~(\ref{eq:defphi}) one has the relation
\begin{eqnarray}
  \widehat{\mathcal{K}}_{\tmop{NZ}} (u) & = & u \mathbbm{1} -
  \hat{\Phi}^{- 1} (u),  \label{eq:mappanz}
\end{eqnarray}
where the hat denotes the Laplace transform, and therefore in matrix
representation
\begin{eqnarray}
  \hat{K}_{\tmop{NZ}} (u) & = & u \mathbbm{1} - \hat{F}^{- 1}
  (u) .  \label{eq:matricenz}
\end{eqnarray}
It immediately appears that, given the time evolution map from the
relationships Eq.~(\ref{eq:matricetcl}) and Eq.~(\ref{eq:matricenz}), one can
directly obtain the generator of the master equation in time-convolutionless
form, or the memory kernel for the Nakajima-Zwanzig form respectively, without
resorting to the evaluation of the whole perturbative series, which of course
leads to the same result {\cite{Vacchini2010b}}. Obviously, given the exact
time evolution one does not need the equations of motion. Nevertheless an
exact comprehensive study as feasible in this case allows, as we shall see, to
point out general features, serving as a guide for phenomenological or
approximate treatments.

Starting from Eq.~(\ref{eq:F}) and Eq.~(\ref{eq:matricetcl}) one obtains for
the model of interest
\begin{widetext}
\begin{eqnarray}
  K_{\tmop{TCL}} (t) & = & \left(\begin{array}{cccc}
    0 & 0 & 0 & 0\\
    0 & \tmop{Re} \left[ \frac{\dot{\gamma} (t)}{\gamma \left( t
    \right)} \right] & \tmop{Im} \left[ \frac{\dot{\gamma} \left( t
    \right)}{\gamma (t)} \right] & 0\\
    0 & - \tmop{Im} \left[ \frac{\dot{\gamma} (t)}{\gamma \left(
    t \right)} \right] & \tmop{Re} \left[ \frac{\dot{\gamma} \left( t
    \right)}{\gamma (t)} \right] & 0\\
    \frac{\left[ 1 - 2 \beta (t) \right] \dot{\alpha} \left( t
    \right) - \left[ 1 - 2 \alpha (t) \right] \dot{\beta} \left(
    t \right)}{\beta (t) + \alpha (t) - 1} & 0 & 0 &
    \frac{\dot{\beta} (t) + \dot{\alpha} (t)}{\beta
    (t) + \alpha (t) - 1}
  \end{array}\right),  \label{eq:ktcl}
\end{eqnarray}
and the expression is well defined provided the determinant
\begin{eqnarray}
  \det F (t) & = & \left| \gamma (t) \right|^2
  \left[ \alpha (t) + \beta (t) - 1 \right] 
  \label{eq:detf}
\end{eqnarray}
is different from zero.

On a similar footing one can consider the Laplace transform of
Eq.~(\ref{eq:F}), given by the matrix $\hat{F} (u)$ with
determinant
\begin{eqnarray}
  \det \hat{F} (u) & = & \left[ \widehat{\gamma_R}^2 \left( u
  \right) + \widehat{\gamma_I}^2 (u) \right] \left[
  \frac{\hat{\alpha} (u) + \hat{\beta} (u)}{u} -
  \frac{1}{u^2} \right],  \label{eq:detfu}
\end{eqnarray}
and using Eq.~(\ref{eq:matricenz}) one further obtains
\begin{eqnarray}
  \hat{K}_{\tmop{NZ}} (u) & = & \left(\begin{array}{cccc}
    0 & 0 & 0 & 0\\
    0 & u - \frac{\widehat{\gamma_R} (u)}{\widehat{\gamma_R}^2 \left( u
    \right) + \widehat{\gamma_I}^2 (u)} & 
    \frac{\widehat{\gamma_I} (u)}{\widehat{\gamma_R}^2 (u) +
    \widehat{\gamma_I}^2 (u)} & 0\\
    0 & - \frac{\widehat{\gamma_I} (u)}{\widehat{\gamma_R}^2 (u)
    + \widehat{\gamma_I}^2 (u)} & u - \frac{\widehat{\gamma_R}
    (u)}{\widehat{\gamma_R}^2 (u) + \widehat{\gamma_I}^2 \left( u
    \right)} & 0\\
    \frac{u^2 \left[ \hat{\alpha} (u) - \hat{\beta} (u) \right]}{1 - u \left[
    \hat{\alpha} (u) + \hat{\beta} (u) \right]} & 0 & 0 & \frac{2 u - u^2
    \left[ \hat{\alpha} (u) + \hat{\beta} (u) \right]}{1 - u \left[
    \hat{\alpha} (u) + \hat{\beta} (u) \right]}
  \end{array}\right),  \label{eq:knz}
\end{eqnarray}
\end{widetext}
which upon inverse Laplace transform provides the exact Nakajima-Zwanzig
integral kernel. As it appears, working with the matrix representation has
proved very convenient to easily obtain the maps fixing the differential and
integrodifferential equations of motion for the model, given by
Eq.~(\ref{eq:mstcl}) and Eq.~(\ref{eq:msnz}) respectively, in terms of the
completely positive time evolution map Eq.~(\ref{eq:rd}).
\subsection{Operator expression}

We now recast the obtained maps, providing time local generator and memory
kernel, in operator form, to better compare with previous work and appreciate
the difference in the obtained expressions. This is easily done observing that
a matrix of the form
\begin{eqnarray}
  A & = & \left(\begin{array}{cccc}
    0 & 0 & 0 & 0\\
    0 & E_r & E_i & 0\\
    0 & - E_i & E_r & 0\\
    X & 0 & 0 & Y
  \end{array}\right)  \label{eq:a}
\end{eqnarray}
in the basis $\left\{ \frac{1}{\sqrt{2}} \mathbbm{1}, \frac{1}{\sqrt{2}}
\sigma_i \right\}$ corresponds in operator form to the map
\begin{eqnarray}
  \mathcal{A} \rho & = & iE_i \left[ \sigma_+ \sigma_-, \rho \right] 
  \label{eq:aop}\\
  &  & + \frac{1}{2} \left( X - Y \right) \left[ \sigma_+ \rho \sigma_- -
  \frac{1}{2} \left\{ \sigma_- \sigma_+, \rho \right\} \right] \nonumber\\
  &  & - \frac{1}{2} \left( X + Y \right) \left[ \sigma_- \rho \sigma_+ -
  \frac{1}{2} \left\{ \sigma_+ \sigma_-, \rho \right\} \right] \nonumber\\
  &  & + \frac{1}{4} \left( Y - 2 E_r \right) \left[ \sigma_z \rho
  \sigma^{}_z - \rho \right], \nonumber
\end{eqnarray}
whose last term can be written in alternative ways according to the identities
\begin{eqnarray}
  \sigma_z \rho \sigma^{}_z - \rho & = & 4 \left[ \sigma_+ \sigma_- \rho
  \sigma_+ \sigma_- - \frac{1}{2} \left\{ \sigma_+ \sigma_-, \rho \right\}
  \right]  \label{eq:equiv}\\
  & = & 4 \left[ \sigma_- \sigma_+ \rho \sigma_- \sigma_+ - \frac{1}{2}
  \left\{ \sigma_- \sigma_+, \rho \right\} \right] . \nonumber
\end{eqnarray}
Exploiting this result one obtains the exact time-convolutionless master
equation describing the reduced dynamics of a two-level atom coupled according
to the Jaynes-Cummings model to a single mode of the radiation field, which is
of the form Eq.~(\ref{eq:mstcl}) with $\mathcal{K}_{\tmop{TCL}} \left( t
\right)$ given by
\begin{multline}
  \mathcal{K}_{\tmop{TCL}} (t) \rho= 
    i \tmop{Im} \left[
  \frac{\dot{\gamma} (t)}{\gamma (t)} \right] \left[
  \sigma_+ \sigma_-, \rho \right]  \label{eq:tcln}\\
   + \frac{\left[ \alpha (t) - 1 \right] \dot{\beta} \left(
  t \right) - \beta (t) \dot{\alpha} (t)}{\beta
  (t) + \alpha (t) - 1}  \left[ \sigma_+ \rho \sigma_- - \frac{1}{2} \left\{ \sigma_-
  \sigma_+, \rho \right\} \right] \\
   + \frac{\left[ \beta (t) - 1 \right] \dot{\alpha} \left(
  t \right) - \alpha (t) \dot{\beta} (t)}{\beta
  (t) + \alpha (t) - 1}  \left[ \sigma_- \rho \sigma_+ - \frac{1}{2} \left\{ \sigma_+
  \sigma_-, \rho \right\} \right]\\
   + \frac{1}{4} \left\{ \frac{\dot{\beta} (t) +
  \dot{\alpha} (t)}{\beta (t) + \alpha \left( t
  \right) - 1} - 2 \tmop{Re} \left[ \frac{\dot{\gamma} \left( t
  \right)}{\gamma (t)} \right] \right\}  \left[ \sigma_z \rho \sigma^{}_z - \rho \right] . 
\end{multline}
In a similar way one has for the Laplace transform of the memory kernel
$\mathcal{K}_{\tmop{NZ}} (t)$, appearing in the exact
Nakajima-Zwanzig master equation Eq.~(\ref{eq:msnz}), the expression
\begin{multline}
  \widehat{\mathcal{K}}_{\tmop{NZ}} (u) \rho  = i
  \frac{\widehat{\gamma_I} (u)}{\widehat{\gamma_R}^2 (u) +
  \widehat{\gamma_I}^2 (u)} \left[ \sigma_+ \sigma_-, \rho
  \right]  \label{eq:nzn}\\
   + \frac{u \left[ u \hat{\alpha} (u) - 1 \right]}{1 - u \left[
  \hat{\alpha} (u) + \hat{\beta} (u) \right]} \left[ \sigma_+ \rho \sigma_- - \frac{1}{2} \left\{ \sigma_-
  \sigma_+, \rho \right\} \right] \\
   + \frac{u \left[ u \hat{\beta} (u) - 1 \right]}{1 - u \left[
  \hat{\alpha} (u) + \hat{\beta} (u) \right]}  \left[ \sigma_- \rho \sigma_+ - \frac{1}{2} \left\{ \sigma_+
  \sigma_-, \rho \right\} \right] \\
   + \frac{1}{4} \left\{ \frac{u^2 \left[ \hat{\alpha} (u) + \hat{\beta}
  (u) \right]}{1 - u \left[ \hat{\alpha} (u) + \hat{\beta} (u) \right]} + 2
  \frac{\widehat{\gamma_R} (u)}{\widehat{\gamma_R}^2 (u) +
  \widehat{\gamma_I}^2 (u)} \right\}\\
   \times \left[ \sigma_z \rho \sigma^{}_z - \rho \right] . 
\end{multline}
Despite being exact these expressions are quite cumbersome, since the
functions given in Eq.~(\ref{eq:abc}), which together with their Laplace
transform determine the structure of these operators, depend on the specific
expression of the environmental state. It is therefore convenient to consider
a specific choice, allowing for a more detailed evaluation.

\subsection{The vacuum case}

If the radiation field is in the vacuum state, the functions given in
Eq.~(\ref{eq:abc}) simplify considerably, since $\alpha (t)
\rightarrow 1$, while $\beta (t)$ becomes a function of $\gamma
(t)$ according to $\beta (t) \rightarrow \left|
\gamma (t) \right|^2$. The function $\gamma (t)$ for
the vacuum case is given by the expression
\begin{eqnarray}
  G^1 (t) & = & e^{i \Delta t / 2}  \left[ \cos \left(
  \frac{\Omega_1 t}{2} \right) - i \frac{\Delta}{\Omega_1} \sin \left(
  \frac{\Omega_1 t}{2} \right) \right],  \label{eq:g1}
\end{eqnarray}
where the superscript recalls that we have a single mode of the radiation
field, while $\Omega_1 = \sqrt{\Delta^2 + 4 g^2}$. These results for the
vacuum case greatly simplify the expression of the obtained master equations,
and inserted in Eq.~(\ref{eq:detf}) show that the time-convolutionless master
equation off-resonance is always well defined.

Indeed the time-convolutionless master equation for the vacuum case simply
reads
\begin{eqnarray}
  \mathcal{K}^{\tmop{Vac}}_{\tmop{TCL}} (t) \rho & = & + i
  \tmop{Im} \left[ \frac{\dot{G^1} (t)}{G^1 (t)}
  \right] \left[ \sigma_+ \sigma_-, \rho \right]  \label{eq:tclvac}\\
  &  & - 2 \tmop{Re} \left[ \frac{\dot{G^1} (t)}{G^1 \left( t
  \right)} \right] \left[ \sigma_- \rho \sigma_+ - \frac{1}{2} \left\{
  \sigma_+ \sigma_-, \rho \right\} \right], \nonumber
\end{eqnarray}
corresponding as expected to a Lindblad structure with time dependent
coefficients. Exploiting Eq.~(\ref{eq:g1}) one has the explicit expression
\begin{multline}
  \mathcal{K}^{\tmop{Vac}}_{\tmop{TCL}} (t) \rho  =\\
  - ig^2
  \Delta \frac{1 - \cos \left( \Omega_1 t \right)}{\Omega_1} \left[ \cos^2 \left( \frac{\Omega_1 t}{2} \right) +
  \frac{\Delta^2 }{\Omega^2_1} \sin^2 \left( \frac{\Omega_1 t}{2} \right)
  \right]^{- 1}  \\
   \times \left[ \sigma_+ \sigma_-, \rho \right] \\
   + 2 g^2 \frac{\sin \left( \Omega_1 t \right)}{\Omega_1} \left[ \cos^2 \left( \frac{\Omega_1 t}{2} \right) +
  \frac{\Delta^2 }{\Omega^2_1} \sin^2 \left( \frac{\Omega_1 t}{2} \right)
  \right]^{- 1} \\
   \times \left[ \sigma_- \rho \sigma_+ - \frac{1}{2} \left\{ \sigma_+
  \sigma_-, \rho \right\} \right],  \label{eq:tclvact}
\end{multline}
where in particular one directly sees that the coefficient in front of the
dissipative term at the r.h.s. of Eq.~(\ref{eq:tclvact}) periodically takes on
negative values, so that it describes a truly non-Markovian behavior and a
Markovian approximation is not justified even for long times.

The choice of the vacuum as bath state brings in important simplifications
also for the expression of the Nakajima-Zwanzig memory kernel, whose Laplace
transform reads
\begin{multline}
  \widehat{\mathcal{K}}^{\tmop{Vac}}_{\tmop{NZ}} (u) \rho  =  +
  i \frac{\widehat{G^1_I} (u)}{\widehat{G^1_R}^2 (u) +
  \widehat{G^1_I}^2 (u)} \left[ \sigma_+ \sigma_-, \rho \right] 
  \label{eq:nzvacu}\\
  + \left[ \frac{1 - u \hat{z}^1 (u)}{\hat{z}^1 \left( u
  \right)} \right]  \left[ \sigma_- \rho \sigma_+ - \frac{1}{2} \left\{ \sigma_+
  \sigma_-, \rho \right\} \right] \\
   - \frac{1}{4} \left[ \frac{1 - u \hat{z}^1 (u)}{\hat{z}^1
  (u)} + 2 \left( u - \frac{\widehat{G^1_R}
  (u)}{\widehat{G^1_R}^2 (u) + \widehat{G^1_I}^2 \left( u
  \right)} \right) \right] \\
   \times  \left[ \sigma_z \rho \sigma^{}_z - \rho \right], 
\end{multline}
where
\begin{eqnarray}
  z^1 (t) & = & \left| G^1 (t) \right|^2 . 
  \label{eq:z1}
\end{eqnarray}
Taking into account the explicit result Eq.~(\ref{eq:g1}) one obtains after
some calculations the following compact expression for the memory kernel
\begin{multline}
  \mathcal{K}^{\tmop{Vac}}_{\tmop{NZ}} \left( \tau \right) \rho  = - ig^2
  \sin \left( \Delta \tau \right) \left[ \sigma_+ \sigma_-, \rho \right] 
  \label{eq:nzvact}\\
   + 2 g^2 \cos \left( \sqrt{\Delta^2 + 2 g^2} \tau \right)  \left[ \sigma_- \rho \sigma_+ - \frac{1}{2} \left\{ \sigma_+
  \sigma_-, \rho \right\} \right] \\- \frac{1}{2} g^2 \left[ \cos \left( \sqrt{\Delta^2 + 2 g^2} \tau
  \right) - \cos \left( \Delta \tau \right) \right]  \left[ \sigma_z \rho \sigma^{}_z - \rho \right], 
\end{multline}
which is always well-defined even on-resonance. These results already allow
for a few important remarks. We first notice that the different operator
contributions in Lindblad form appearing in the various time local and
integral kernels are multiplied by different time dependent functions. This
lesson, already learnt in other models
{\cite{Salo2006a,Breuer2008a,Cresser2010a,Vacchini2010b}}, tells us that as a
general rule non-Markovian dynamics cannot be obtained from Markovian
expressions by simply taking the convolution with a single integral kernel, or
multiplying by a single time dependent function. This fact has often been
overlooked when seeking non-Markovian integrodifferential master equations
{\cite{Barnett2001a,Shabani2005a,Maniscalco2007a,Wilkie2009a}}, sometimes
leading to unphysical behaviors. More than this, for the same model different
sets of non-Markovian equations of motion can have different operator
structures, as it appears comparing e.g. the time-convolutionless and
Nakajima-Zwanzig results for the vacuum Eq.~(\ref{eq:tclvac}) and
Eq.~(\ref{eq:nzvact}). This fact has already been noticed in
{\cite{Vacchini2010b}}, but the present analysis shows that this asymmetry
depends on the choice of environmental state. For the present model it only
appears in connection with the vacuum state. Indeed, while the disappearance
of the term corresponding to excitation of the two-level system is obvious on
physical grounds, when considering as bath state the vacuum, the vanishing of
the coefficient in front of the dephasing term $\sigma_z \rho \sigma^{}_z -
\rho$ is a peculiar feature of the time-convolutionless master equation.

\section{Damped two-level system\label{sec:djc}}

\subsection{Exact master equations for bath in the vacuum state}

The technique used in Sect.~\ref{sec:ms} to obtain the time-convolutionless
and Nakajima-Zwanzig equation of motions for a model whose evolution is known,
by exploiting the representation of maps in terms of matrices, is applicable
for a detailed study of simple models. We exploit it now for a variant of the
model considered in Sect.~\ref{sec:jc}, in which the environmental Hamiltonian
is a collection of harmonic oscillators
\begin{eqnarray}
  H_E & = & \sum_k \bignone \omega_k b_k^{\dag} b_k,  \label{eq:hen}
\end{eqnarray}
and the interaction Hamiltonian is replaced by
\begin{eqnarray}
  H_I & = & \sum_k \bignone \left( g_k \sigma_+ \otimes b_k + g_k^{\ast}
  \sigma_- \otimes b_k^{\dag} \right) .  \label{eq:hin}
\end{eqnarray}
This model corresponds for a Lorentzian spectral density to the damped
Jaynes-Cummings model. The time evolution map for this model, considering the
special case of an environment in the vacuum state, has been obtained in
{\cite{Breuer1999b}} and can be expressed as:
\begin{eqnarray}
  \rho_{11} (t) & = & \rho_{11} \left( 0 \right) \left| G \left(
  t \right) \right|^2, \nonumber\\
  \rho_{10} (t) & = & \rho_{10} \left( 0 \right) G \left( t
  \right), \nonumber\\
  \rho_{01} (t) & = & \rho_{01} \left( 0 \right) G^{\ast} \left(
  t \right), \nonumber\\
  \rho_{00} (t) & = & \rho_{00} \left( 0 \right) + \rho_{11}
  \left( 0 \right) \left( 1 - \left| G (t) \right|^2 \right), 
  \label{eq:rdd}
\end{eqnarray}
where $\rho (t) = \Phi^{\tmop{DVac}} (t) \rho \left(
0 \right)$, since we are considering the damped
model with the bath in the vacuum state. The function $G (t)$ is the solution of the equation
\begin{eqnarray}
  \frac{d}{dt} G (t) = - \int^t_0 \tmop{dt}_1 f \left( t - t_1
  \right) G \left( t_1 \right) &  & G \left( 0 \right) = 1,  \label{eq:G}
\end{eqnarray}
with $f (t)$ the two-point correlation function given by
\begin{eqnarray}
  f \left( t - t_1 \right) & = & e^{i \omega_0 \left( t - t_1 \right)} \langle
  0| \sum_k \bignone g_k b_k e^{- i \omega_k t} \sum_j \bignone g^{\ast}_j
  b^{\dag}_j e^{i \omega_j t_1} |0 \rangle \bignone \nonumber\\
  & = & \sum_k \left| g_k \right|^2 e^{i \left( \omega_0 - \omega_k \right)
  \left( t - t_1 \right)},  \label{eq:f}
\end{eqnarray}
corresponding to the Fourier transform of the spectral density.

Starting from Eq.~(\ref{eq:rdd}) and exploiting the same strategy used in
Sect.~\ref{sec:ms} one immediately obtains for the matrix representation of
the time-convolutionless generator the expression
\begin{widetext}
\begin{eqnarray}
  K^{\tmop{DVac}}_{\tmop{TCL}} (t) & = &
  \left(\begin{array}{cccc}
    0 & 0 & 0 & 0\\
    0 & \tmop{Re} \left[ \frac{\dot{G} (t)}{G (t)}
    \right] & \tmop{Im} \left[ \frac{\dot{G} (t)}{G \left( t
    \right)} \right] & 0\\
    0 & - \tmop{Im} \left[ \frac{\dot{G} (t)}{G (t)}
    \right] & \tmop{Re} \left[ \frac{\dot{G} (t)}{G \left( t
    \right)} \right] & 0\\
    2 \tmop{Re} \left[ \frac{\dot{G} (t)}{G (t)}
    \right] & 0 & 0 & 2 \tmop{Re} \left[ \frac{\dot{G} (t)}{G
    (t)} \right]
  \end{array}\right),  \label{eq:tcld}
\end{eqnarray}
so that the master equation in operator form reads
\begin{eqnarray}
  \mathcal{K}^{\tmop{DVac}}_{\tmop{TCL}} (t) \rho & = & + i
  \tmop{Im} \left[ \frac{\dot{G^{}} (t)}{G (t)}
  \right] \left[ \sigma_+ \sigma_-, \rho \right]  \label{eq:tclvacd}\\
  &  & - 2 \tmop{Re} \left[ \frac{\dot{G} (t)}{G \left( t
  \right)} \right] \left[ \sigma_- \rho \sigma_+ - \frac{1}{2} \left\{
  \sigma_+ \sigma_-, \rho \right\} \right], \nonumber
\end{eqnarray}
which confirms the result obtained in {\cite{Breuer2007}}. One can also
determine the expression of the Nakajima-Zwanzig memory kernel, whose Laplace
transform is given by
\begin{eqnarray}
  \hat{K}^{\tmop{DVac}}_{\tmop{NZ}} (u) & = &
  \left(\begin{array}{cccc}
    0 & 0 & 0 & 0\\
    0 & u - \frac{\widehat{G_R} (u)}{\widehat{G_R}^2 (u) + \widehat{G_I}^2
    (u)} &  \frac{\widehat{G_I} (u)}{\widehat{G_R}^2 (u) + \widehat{G_I}^2
    (u)} & 0\\
    0 & - \frac{\widehat{G_I} (u)}{\widehat{G_R}^2 (u) + \widehat{G_I}^2 (u)}
    \hspace*{\fill} & u - \frac{\widehat{G_R} (u)}{\widehat{G_R}^2 (u) +
    \widehat{G_I}^2 (u)} & 0\\
  \frac{u \hat{z} (u)-1}{\hat{z} (u)} & 0 & 0 & \frac{u \hat{z} (u)-1}{\hat{z} (u)}
  \end{array}\right),  \label{eq:nzd}
\end{eqnarray}
\end{widetext}
where we have used the notation
\begin{eqnarray}
  z^{} (t) & = & \left| G (t) \right|^2, 
  \label{eq:z}
\end{eqnarray}
leading to the master equation
\begin{multline}
  \widehat{\mathcal{K}}^{\tmop{Vac}}_{\tmop{NZ}} (u) \rho  =  +
  i \frac{\widehat{G_I} (u)}{\widehat{G^{}_R}^2 (u) +
  \widehat{G^{}_I}^2 (u)} \left[ \sigma_+ \sigma_-, \rho \right] 
  \label{eq:nzvacdu}\\
  + \left[ \frac{1 - u \hat{z}^{} (u)}{\hat{z}^{} \left( u
  \right)} \right] \left[ \sigma_- \rho \sigma_+ - \frac{1}{2} \left\{ \sigma_+
  \sigma_-, \rho \right\} \right]\\
   - \frac{1}{4} \left[ \frac{1 - u \hat{z}^{} \left( u
  \right)}{\hat{z}^{} (u)} + 2 \left( u - \frac{\widehat{G^{}_R}
  (u)}{\widehat{G^{}_R}^2 (u) + \widehat{G^{}_I}^2 \left( u
  \right)} \right) \right] \\ \times \left[ \sigma_z \rho \sigma^{}_z - \rho \right] . 
\end{multline}

\subsection{Role of correlation function}

We now show that this constructive approach leads to the same result as
obtained by a suitable ansatz in {\cite{Vacchini2010b}}, where the detailed
connection with the perturbative expansion via projection operator techniques
has also been considered. We first have to recall the relation in Laplace
transform between the functions $G (t)$ and $f (t)$.
Indeed, for the case of a two-level system interacting with an environment of
oscillators in the vacuum state, all functions appearing in the non-Markovian
equations of motion can be related to a single correlation function of the
model, given by Eq.~(\ref{eq:f}).

To fully exploit this fact in order to express the memory kernel in the most
compact and transparent way let us observe that according to Eq.~(\ref{eq:G})
one has
\begin{equation}
  \hat{f}  (u) = \frac{\hat{G}^{\ast} (u)}{\left|
  \widehat{G^{}_{}} (u) \right|^2} - u,  \label{eq:gf}
\end{equation}
so that in view of the fact that for real $u$
\begin{eqnarray}
  \tmop{Re} \left[ \hat{h} (u) \right] & = & \widehat{\tmop{Re}
  h} (u)  \label{eq:re}
\end{eqnarray}
together with
\begin{eqnarray}
  \tmop{Im} \left[ \hat{h} (u) \right] & = & \widehat{\tmop{Im}
  h} (u),  \label{eq:im}
\end{eqnarray}
the functions
\begin{eqnarray}
  \widehat{f_I} (u) & \tmop{and} & - \frac{\widehat{G_I}
  (u)}{\widehat{G_R}^2 (u) + \widehat{G_I}^2 (u)}  \label{eq:fi}
\end{eqnarray}
do coincide on the real axis, and therefore due to the identity principle on
the common region of analyticity. A corresponding result holds for the
functions
\begin{eqnarray}
  \widehat{f_R} (u) & \tmop{and} & \frac{\widehat{G_R}
  (u)}{\widehat{G_R}^2 (u) + \widehat{G_I}^2 (u)} - u,  \label{eq:fr}
\end{eqnarray}
so that the matrix representing the Nakajima-Zwanzig kernel can be rewritten
in a more compact way as
\begin{equation}
  \hat{K}^{\tmop{DVac}}_{\tmop{NZ}} (u) =
  \left(\begin{array}{cccc}
    0 & 0 & 0 & 0\\
    0 & -\widehat{f_R} (u) & - \widehat{f_I} (u) & 0\\
    0 & \widehat{f_I} (u) \hspace*{\fill} & -\widehat{f_R} \left(
    u \right) & 0\\
   \frac{u \hat{z} (u)-1}{\hat{z} (u)} & 0 & 0 & \frac{u \hat{z} (u)-1}{\hat{z} (u)}
 \end{array}\right),  \label{eq:nzdf}
\end{equation}
leading indeed to the integral kernel first obtained in {\cite{Vacchini2010b}}
\begin{eqnarray}
  \mathcal{K}^{\tmop{DVac}}_{\tmop{NZ}} \left( \tau \right) \rho & = & - if_I
  \left( \tau \right) \left[ \sigma_+ \sigma_-, \rho \right] 
  \label{eq:kernel2}\\
  &  & + k_1 \left( \tau \right) \left[ \sigma_- \rho \sigma_+ - \frac{1}{2}
  \left\{ \sigma_+ \sigma_-, \rho \right\} \right] \nonumber\\
  &  & - \frac{1}{4} \left[ k_1 \left( \tau \right) - 2 f_R \left( \tau
  \right) \right] \left[ \sigma_z \rho \sigma^{}_z - \rho \right], \nonumber
\end{eqnarray}
where we have set
\begin{eqnarray}
  \hat{k}_1 (u) & = & \frac{1 - u \hat{z} \left( u
  \right)}{\hat{z} (u)} .  \label{eq:kla}
\end{eqnarray}
For the case of a single mode the correlation function considered in
Eq.~(\ref{eq:f}) explicitly becomes
\begin{eqnarray}
  f^1 (t) & = & g^2 \tmmathbf{e}^{i \Delta t},  \label{eq:f1}
\end{eqnarray}
where the superscript again stresses the fact that a single mode is
considered. The solution of the integrodifferential Eq.~(\ref{eq:G}) is then
exactly given by the function $G^1 (t)$ introduced in
Eq.~(\ref{eq:g1}). As it should be, the equations Eq.~(\ref{eq:tclvac}) and
Eq.~(\ref{eq:nzvacu}) are obtained from Eq.~(\ref{eq:tclvacd}) and
Eq.~(\ref{eq:nzvacdu}) under the replacement $G (t) \rightarrow
G^1 (t)$, which corresponds to the special choice of a single
mode bath. At the same time it is clear that the explicit coefficients
appearing in Eq.~(\ref{eq:nzvact}) can indeed be expressed using real and
imaginary part of Eq.~(\ref{eq:f1}), as well as the inverse Laplace transform
of
\begin{eqnarray}
  \hat{k}^1_1 (u) & = & \frac{1 - u \hat{z}^1 \left( u
  \right)}{\hat{z}^1 (u)}  \label{eq:k1u}\\
  & = & 2 g^2 \frac{u}{u^2 + \Delta^2 + 2 g^2}, \nonumber
\end{eqnarray}
which is given by
\begin{eqnarray}
  k^1_1 \left( \tau \right) & = & 2 g^2 \cos \left( \sqrt{\Delta^2 + 2 g^2}
  \tau \right) .  \label{eq:k1t}
\end{eqnarray}
\begin{eqnarray}
  &  &  \nonumber
\end{eqnarray}

\subsection{Additional term for thermal bath}

It is to be stressed, that the possibility to express all relevant functions
appearing in time-convolutionless and Nakajima-Zwanzig master equations with
reference to the single correlation function $f (t)$ is a special
feature of the two-level system coupled to the vacuum. Indeed while the
results of Sect.~\ref{sec:ms} for the vacuum are a special case of the model
considered in Sect.~\ref{sec:djc}, the more general situation of a thermal
bath can be explicitly considered for the case of the reduced dynamics of the
Jaynes-Cummings model, showing that the expectation values of various
operators as in Eq.~(\ref{eq:abc}) have to be specified in order to give the
exact equations of motion.

In Sect.~\ref{sec:ms} we have seen that the time-convolutionless generator
has a different operator structure with respect to the Nakajima-Zwanzig memory
kernel only for the case of the vacuum, as it appears comparing
Eq.~(\ref{eq:tclvac}) and Eq.~(\ref{eq:nzvact}), while this is no more true
for a thermal state. This strongly suggests that the asymmetry in the operator
structure of Eq.~(\ref{eq:tclvacd}) and Eq.~(\ref{eq:kernel2}) is also due to
this special choice of the bath state. To check this fact, in the absence of
the exact time evolution map, one has to calculate the time-convolutionless
master equation relying on the standard perturbative technique
{\cite{Breuer2007}}, considering terms up to fourth order. The necessity to go
up to the fourth perturbative order is immediately clear looking at the
interaction Hamiltonian Eq.~(\ref{eq:hin}), and observing that the dephasing
term, as it appears from Eq.~(\ref{eq:equiv}), involves a quadrilinear
contribution in the raising and lowering operators $\sigma_+$ and $\sigma_-$.
This task has been accomplished in Appendix~\ref{sec:appa}, leading to the
result
\begin{eqnarray}
  \mathcal{K}^D_{\tmop{TCL}} (t) \rho & = & i \tmop{Im} \gamma_s
  (t) \left[ \sigma_+ \sigma_-, \rho \right]  \label{eq:tclnth}\\
  &  & + \gamma_+ (t) \left[ \sigma_+ \rho \sigma_- -
  \frac{1}{2} \left\{ \sigma_- \sigma_+, \rho \right\} \right] \nonumber\\
  &  & + \gamma_- (t) \left[ \sigma_- \rho \sigma_+ -
  \frac{1}{2} \left\{ \sigma_+ \sigma_-, \rho \right\} \right] \nonumber\\
  &  & + \frac{1}{4} \gamma_d (t) \left[ \sigma_z \rho
  \sigma^{}_z - \rho \right], \nonumber
\end{eqnarray}
and the detailed expression of the various coefficients in terms of two- and
four-points correlation functions of the system can be found in
Eq.~(\ref{eq:gamma}) of Appendix~\ref{sec:appa}. \

This result shows that indeed the disappearance of the dephasing term in the
time-convolutionless master equation for the vacuum is a very special feature
of this choice of bath. Once again the operator structure of non-Markovian
equations of motion is strongly dependent on the details of both bath and
interaction term.

\section{Conclusions}
\label{sec:ceo}

We have obtained the exact time-convolutionless, time-local master equation
and Nakajima-Zwanzig integrodifferential master equation for a two-level
system coupled to a single bosonic mode, considering a generic bath state.
This has been possible thanks to the knowledge of the exact dynamics, and
corresponds to the result which can be obtained resumming all terms in the
corresponding perturbative techniques. The path followed here, to associate
matrices to maps and to consider the direct relation between time evolution
map and time local generator, as well as memory kernel, is however much more
straightforward. The result shows that in a realistic model each operator
contribution in Lindblad form has its own time dependent function, responsible
for the non-Markovian behavior, so that a simple multiplication or convolution
of the Markovian result with a single phenomenologically guessed function will
generally not work. Furthermore the operator structures of the equations of
motion in the two cases can strongly differ, also depending on the state of
the bath. In particular it has been shown that a dephasing term, which is
always present in the Nakajima-Zwanzig equations of motion, disappears for the
time-convolutionless case for a bath in the vacuum state. This has been
checked for the undamped two-level system thanks to the exact solution, and
for the damped case by calculating the fourth order contribution to the
time-convolutionless perturbation expansion.

The pursued approach is quite straightforward, even though to obtain the exact
expressions it relies on the knowledge of the full time evolution, which can
obviously only be feasible for exceptional cases. The detailed analysis of
such cases however proves quite useful in understanding the basic features of
a non-Markovian description, in particular it puts into evidence the strict
relationship between the different quantities which appear in the evolution
equations, showing that phenomenological ansatz are in general not easily
feasible. Furthermore, the connection to the microscopic perturbative
derivation techniques as outlined here, can help in determining the operator
structure of the dynamical equations. This structure can sometimes be unveiled
calculating the first perturbative contributions, thus restricting the number
of ansatz necessary in order to consider a sound phenomenological model.

The necessity to consider non-Markovian equations of motion is often due to
strong coupling effects between system and reservoir. This in turn implies
that a factorized initial state is not always the most appropriate initial
condition, so that the study of the dynamics of correlated initial states
becomes of great significance. This long standing topic has recently seen
important advancements
{\cite{Pechukas1994a,Alicki1995a,Royer1996a,vanWonderen2006a,Shabani2009a,Rodriguez-Rosario2010a,Aniello-xxx,Laine-xxx}},
and we plan to address it in future research work, looking in simple but
realistic models, such as the one considered in this article, for general
signatures of the effect of initial correlations. These results might help to
consider, by means of some approximation, more general systems.

\section{Acknowledgments}

The authors are grateful to Heinz-Peter Breuer, Federico Casagrande and
Ludovico Lanz for helpful discussions and reading of the manuscript. This work
was partially supported by Ministero dell'Istruzione, dell'Universit\`a e
della Ricerca (MIUR) under PRIN2008.
\begin{widetext}

\appendix
\section{}\label{sec:appa}

We here consider the contributions up to fourth order of the
time-convolutionless projection operator technique for the damped two-level
system considered in Sect.~\ref{sec:djc}. We assume for the environment an
equilibrium state $\rho_E$, which commutes with the number operator, and we
use the standard projection operator
\begin{eqnarray}
  \mathcal{P} w & = & \tmop{Tr}_E \left( w \right) \otimes \rho_E, 
  \label{eq:proj}
\end{eqnarray}
where $w$ denotes a state of system and environment.

The time-convolutionless perturbative expansion can be obtained by repeated
action of the projection operator considered above and of the superoperator
\begin{eqnarray}
  \mathcal{L} (t) w & = & - i \left[ H_I (t), w
  \right],  \label{eq:l}
\end{eqnarray}
where $H_I (t)$ is the interaction picture expression
corresponding to Eq.~(\ref{eq:hin}), given by
\begin{eqnarray}
  H_I (t) & = & \sigma_+ (t) \otimes B \left( t
  \right) + \sigma_- (t) \otimes B^{\dag} (t), 
  \label{eq:hint}
\end{eqnarray}
with
\begin{eqnarray}
  \sigma_{\pm} (t) & = & e^{\pm i \omega_0 t}  \label{eq:sigmat}
\end{eqnarray}
and
\begin{eqnarray}
  B (t) & = & \sum_k \bignone g_k b_k e^{- i \omega_k t} . 
  \label{eq:bt}
\end{eqnarray}
The contributions to second- and fourth-order for the time-convolutionless
generator then read {\cite{Breuer2007}}
\begin{eqnarray}
  \mathcal{K}_2 (t) & = & \int^t_0 d t_1 \mathcal{P} \mathcal{L}
  (t) \mathcal{L} (t_1) \mathcal{P}  \label{eq:k2}
\end{eqnarray}
and
\begin{eqnarray}
  \mathcal{K}_4 (t) & = & \int^t_0 d t_1 \int^{t_1}_0 d t_2
  \int^{t_2}_0 d t_3 \left[ \mathcal{P} \mathcal{L} (t) \mathcal{L} (t_1)
  \mathcal{L} (t_2) \mathcal{L} (t_3) \mathcal{P} - \mathcal{P} \mathcal{L}
  (t) \mathcal{L} (t_1) \mathcal{P} \mathcal{L} (t_2) \mathcal{L} (t_3)
  \mathcal{P}  \label{eq:k4} \right.\\
  &  & \left. - \mathcal{P} \mathcal{L} (t) \mathcal{L} (t_2) \mathcal{P}
  \mathcal{L} (t_1) \mathcal{L} (t_3) \mathcal{P} - \mathcal{P} \mathcal{L}
  (t) \mathcal{L} (t_3) \mathcal{P} \mathcal{L} (t_1) \mathcal{L} (t_2)
  \mathcal{P} \right] \nonumber
\end{eqnarray}
respectively. Using these expressions one immediately obtains the operator
form of the contributions to the time-convolutionless master equation for the
reduced dynamics according to
\begin{eqnarray}
  \mathcal{K}^{\left( 2 \right)}_{\tmop{TCL}} (t) \rho & = &
  \tmop{Tr}_E \left\{ \int^t_0 d t_1 \mathcal{L} (t) \mathcal{L} (t_1)
  \mathcal{\rho} \otimes \rho_E \right\}  \label{eq:tcl22}
\end{eqnarray}
and
\begin{eqnarray}
  \mathcal{K}^{\left( 4 \right)}_{\tmop{TCL}} (t) \rho & = &
  \tmop{Tr}_E \left\{\int^t_0 d t_1 \int^{t_1}_0 d t_2 \int^{t_2}_0 d t_3 
  \label{eq:tcl44} \right.\\
  &  & \left[ \mathcal{P} \mathcal{L} (t) \mathcal{L} (t_1) \mathcal{L} (t_2)
  \mathcal{L} (t_3) \mathcal{P} \mathcal{\rho} \otimes \rho_E - \mathcal{P}
  \mathcal{L} (t) \mathcal{L} (t_1) \mathcal{P} \mathcal{L} (t_2) \mathcal{L}
  (t_3) \mathcal{P} \mathcal{\rho} \otimes \rho_E \right. \nonumber\\
  &  & \left. \left. - \mathcal{P} \mathcal{L} (t) \mathcal{L} (t_2) \mathcal{P}
  \mathcal{L} (t_1) \mathcal{L} (t_3) \mathcal{P} \rho \otimes \rho_E -
  \mathcal{P} \mathcal{L} (t) \mathcal{L} (t_3) \mathcal{P} \mathcal{L} (t_1)
  \mathcal{L} (t_2) \mathcal{P} \rho \otimes \rho_E \right]\right\}  . \nonumber
\end{eqnarray}
We recall that such high order contributions are needed in order to check the
appearance, for an environmental state different from the vacuum, of the
dephasing term $\sigma_z \rho \sigma_z - \rho$, which involves expressions
with altogether four raising and lowering operators of the two-level system.

The second order contribution Eq.~(\ref{eq:tcl22}) can be expressed by means
of the following two correlation functions:
\begin{eqnarray}
  f (t - t_1) & = & e^{i \omega_0 (t_{} - t_1)} \tmop{Tr}_E \left\{ B (t)
  B^{\dag} (t_1) \rho_E \right\}  \label{eq:fe}\\
  & = & \sum_k \left| g_k \right|^2 e^{i \left( \omega_0 - \omega_k \right)
  \left( t - t_1 \right)} \langle n_k + 1 \rangle_E, \nonumber
\end{eqnarray}
which corresponds to Eq.~(\ref{eq:f}) if the bath is in the vacuum state, and
\begin{eqnarray}
  g (t - t_1) & = & e^{- i \omega_0 (t_{} - t_1)} \tmop{Tr}_E \left\{ B^{\dag}
  (t) B (t_1) \rho_E \right\}  \label{eq:ge}\\
  & = & \sum_k \left| g_k \right|^2 e^{- i \left( \omega_0 - \omega_k \right)
  \left( t - t_1 \right)} \langle n_k \rangle_E, \nonumber
\end{eqnarray}
which vanishes in the vacuum. In terms of these functions one has
\begin{eqnarray}
  \mathcal{P} \mathcal{L} (t_{\alpha}) \mathcal{L} (t_{\beta}) \mathcal{P}
  \mathcal{\rho} \otimes \rho_E & = & - \left[ f (t_{\alpha} - t_{\beta})
  \sigma_+ \sigma_- \rho + f^{\ast} (t_{\alpha} - t_{\beta}) \rho \sigma_+
  \sigma_-  \label{eq:pp} \right.\\
  &  & + g (t_{\alpha} - t_{\beta}) \sigma_- \sigma_+ \rho + g^{\ast}
  (t_{\alpha} - t_{\beta}) \rho \sigma_- \sigma_+ \nonumber\\
  &  & \left. - 2 \tmop{Re} f (t_{\alpha} - t_{\beta}) \sigma_- \rho \sigma_+
  - 2 \tmop{Re} g (t_{\alpha} - t_{\beta}) \sigma_+ \rho \sigma_- \right]
  \otimes \rho_E . \nonumber
\end{eqnarray}
This result is sufficient to obtain the time-convolutionless master equation
up to second order, indeed upon inserting Eq.~(\ref{eq:pp}) in
Eq.~(\ref{eq:tcl22}) one obtains
\begin{eqnarray}
  \mathcal{K}^{\left( 2 \right)}_{\tmop{TCL}} (t) \rho & = & - i
  \left[ \mathfrak{f}_I (t) +\mathfrak{g}_I (t)
  \right]  \left[ \sigma_+ \sigma_-, \rho \right]  \label{eq:th2}\\
  &  & + 2\mathfrak{f}_R (t) \left[ \sigma_+ \rho \sigma_- -
  \frac{1}{2} \left\{ \sigma_- \sigma_+, \rho \right\} \right] \nonumber\\
  &  & + 2\mathfrak{g}_R (t) \left[ \sigma_- \rho \sigma_+ -
  \frac{1}{2} \left\{ \sigma_+ \sigma_-, \rho \right\} \right], \nonumber
\end{eqnarray}
where we have set
\begin{eqnarray}
  \mathfrak{f} (t) & = & \int^t_0 d t_1 f (t - t_1) 
  \label{eq:ff}
\end{eqnarray}
and
\begin{eqnarray}
  \mathfrak{g} (t) & = & \int^t_0 d t_1 g (t - t_1), 
  \label{eq:gg}
\end{eqnarray}
denoting as usual real and imaginary parts with the subscripts $R$ and $I$
respectively.

To consider the fourth order contribution one has to evaluate the four terms
given in Eq.~(\ref{eq:tcl44}). The last three terms at the right hand side can
be obtained applying twice the result Eq.~(\ref{eq:pp}), thus obtaining
\begin{eqnarray}
  \mathcal{P} \mathcal{L} (t_{}) \mathcal{L} (t_{\alpha}) \mathcal{P}
  \mathcal{} \mathcal{L} (t_{\beta}) \mathcal{L} (t_{\gamma}) \mathcal{P} \rho
  \otimes \rho_E & = & \left[ - 4 \sigma_+ \rho \sigma_- \left\{ \tmop{Re} f
  (t - t_{\alpha}) \tmop{Re} g (t_{\beta} - t_{\gamma}) + \tmop{Re} g (t -
  t_{\alpha}) \tmop{Re} g (t_{\beta} - t_{\gamma}) \right\}  \label{eq:pppp}
  \right.\\
  &  & - 4 \sigma_- \rho \sigma_+ \left\{ \tmop{Re} g (t - t_{\alpha})
  \tmop{Re} f (t_{\beta} - t_{\gamma}) + \tmop{Re} f (t - t_{\alpha})
  \tmop{Re} f (t_{\beta} - t_{\gamma}) \right\} \nonumber\\
  &  & + \sigma_+ \sigma_- \rho_{} f (t_{} - t_{\alpha}) f (t_{\beta} -
  t_{\gamma}) + \rho \sigma_+ \sigma_- f^{\ast} (t_{} - t_{\alpha}) f^{\ast}
  (t_{\beta} - t_{\gamma}) \nonumber\\
  &  & + \sigma_- \sigma_+ \rho_{} g (t_{} - t_{\alpha}) g (t_{\beta} -
  t_{\gamma}) + \rho \sigma_- \sigma_+ g^{\ast} (t_{} - t_{\alpha}) g^{\ast}
  (t_{\beta} - t_{\gamma}) \nonumber\\
  &  & + \sigma_+ \sigma_- \rho_{} \sigma_+ \sigma_- \left\{ 2 \tmop{Re}
  \left[ f (t_{} - t_{\alpha}) f^{\ast} (t_{\beta} - t_{\gamma}) \right]
  \right. \nonumber\\
  &  & \left. + 4 \tmop{Re} g (t_{} - t_{\alpha}) \tmop{Re} f (t_{\beta} -
  t_{\gamma}) \right\} \nonumber\\
  &  & + \sigma_- \sigma_+ \rho_{} \sigma_- \sigma_+ \left\{ 2 \tmop{Re}
  \left[ g (t_{} - t_{\alpha}) g^{\ast} (t_{\beta} - t_{\gamma}) \right]
  \right. \nonumber\\
  &  & \left. + 4 \tmop{Re} f (t_{} - t_{\alpha}) \tmop{Re} g (t_{\beta} -
  t_{\gamma}) \right\} \nonumber\\
  &  & + \sigma_- \sigma_+ \rho \sigma_+ \sigma_- \left\{ f^{\ast} (t_{} -
  t_{\alpha}) g (t_{\beta} - t_{\gamma}) + g (t - t_{\alpha}) f^{\ast}
  (t_{\beta} - t_{\gamma}) \right\} \nonumber\\
  &  & \left. + \sigma_+ \sigma_- \rho \sigma_- \sigma_+ \left\{ g^{\ast} (t
  - t_{\alpha}) f (t_{\beta} - t_{\gamma}) + f (t - t_{\alpha}) g^{\ast}
  (t_{\beta} - t_{\gamma}) \right\} \right] \otimes \rho_E, \nonumber
\end{eqnarray}
where the relations $\sigma_+^2 = \sigma_-^2 = 0$ have been repeatedly used,
together with the assumption $\left[ \rho_E, n_k \right] = 0$.

The first term at the right hand side of Eq.~(\ref{eq:tcl44}) instead requires
the introduction of a four-point correlation function, which is given by
\begin{eqnarray}
  h (t_a, t_b, t_c, t_d) & = & e^{i \omega_0 (t_a - t_b + t_c - t_d)}
  \tmop{Tr}_E \left\{ B (t_a) B^{\dag} (t_b) B (t_c) B^{\dag} (t_d) \rho_E
  \right\},  \label{eq:hkw}
\end{eqnarray}
with $B^{} (t)$ as in Eq.~(\ref{eq:bt}). An explicit evaluation of
$\mathcal{P} \mathcal{L} (t) \mathcal{L} (t_1) \mathcal{L} (t_2) \mathcal{L}
(t_3) \mathcal{P} \mathcal{\rho} \otimes \rho_E$ together with the repeated
use of Eq.~(\ref{eq:pppp}) then leads to the desired result, which can be
obtained with a straightforward though very lengthy calculation. The fourth
order contribution reads
\begin{eqnarray}
  \mathcal{K}^{\left( 4 \right)}_{\tmop{TCL}} (t) \rho & = & i
  \left[ \mathfrak{p}_I (t) +\mathfrak{r}_I (t)
  +\mathfrak{v}_I (t) \right] \left[ \sigma_+ \sigma_-, \rho
  \right]  \label{eq:th4}\\
  &  & +\mathfrak{t} (t) \left[ \sigma_+ \rho \sigma_- -
  \frac{1}{2} \left\{ \sigma_- \sigma_+, \rho \right\} \right] \nonumber\\
  &  & +\mathfrak{u} (t) \left[ \sigma_- \rho \sigma_+ -
  \frac{1}{2} \left\{ \sigma_+ \sigma_-, \rho \right\} \right] \nonumber\\
  &  & + \frac{1}{4} \left[ \mathfrak{q} (t) +\mathfrak{s}
  (t) + 2\mathfrak{v}_R (t) \right] \left[ \sigma_z
  \rho \sigma^{}_z - \rho \right], \nonumber
\end{eqnarray}
where in analogy to the notation of Eq.~(\ref{eq:ff}) and Eq.~(\ref{eq:gg}) we
use the Fraktur character to denote the triple integral over time of the
function with the corresponding Roman letter, for example
\begin{eqnarray}
  \mathfrak{p} (t) & = & \int^t_0 d t_1 \int^{t_1}_0 d t_2
  \int^{t_2}_0 d t_3 p (t, t_1, t_2, t_3) .  \label{eq:pfrak}
\end{eqnarray}
The functions determining the coefficients appearing in Eq.~(\ref{eq:th4}) are
given in terms of the above introduced two- and four-points correlation
functions of the model according to the expressions
\begin{eqnarray}
  p (t, t_1, t_2, t_3) & = & _{} - \sum_{\alpha \beta \gamma} f (t_{} -
  t_{\alpha}) f (t_{\beta} - t_{\gamma}) + h (t, t_1, t_2, t_3) 
  \label{eq:coeff}\\
  q (t, t_1, t_2, t_3) & = & - 2 \sum_{\alpha \beta \gamma} \left\{ \tmop{Re}
  \left[ f (t_{} - t_{\alpha}) f^{\ast} (t_{\beta} - t_{\gamma}) \right] + 2
  \tmop{Re} g (t_{} - t_{\alpha}) \tmop{Re} f (t_{\beta} - t_{\gamma}) -
  \tmop{Re} h (t_{\alpha}, t_{}, t_{\beta}, t_{\gamma}) \right\} \nonumber\\
  r (t, t_1, t_2, t_3) & = & g (t - t_2) g (t_1 - t_3) + g (t - t_3) g (t_1 -
  t_2) + f (t_1 - t_{}) f (t_3 - t_2) - h (t_1, t_{}, t_3, t_2) \nonumber\\
  s (t, t_1, t_2, t_3) & = & - 2 \sum_{\alpha \beta \gamma} \left\{ \tmop{Re}
  [f (t_{} - t_{\alpha}) f (t_{\gamma} - t_{\beta})] + 2 \tmop{Re} f (t -
  t_{\alpha}) \tmop{Re} g (t_{\beta} - t_{\gamma}) - \tmop{Re} h (t_{},
  t_{\alpha}, t_{\gamma}, t_{\beta}) \right\} \nonumber\\
  t (t, t_1, t_2, t_3) & = & 2 \sum_{\alpha \beta \gamma} \left\{ \tmop{Re} [f
  (t_{} - t_{\alpha}) f (t_{\gamma} - t_{\beta})] + \tmop{Re} [g (t -
  t_{\alpha}) g (t_{\beta} - t_{\gamma})] \right. \nonumber\\
  &  & \left. + 2 \tmop{Re} f (t - t_{\alpha}) \tmop{Re} g (t_{\beta} -
  t_{\gamma}) - \tmop{Re} h (t_{}, t_{\alpha}, t_{\gamma}, t_{\beta}) \right\}
  \nonumber\\
  &  & + 2 \left\{ \tmop{Re} [f (t_1 - t_{}) f (t_3 - t_2)] - \tmop{Re} [g
  (t_{} - t_1) g (t_2 - t_3)] - \tmop{Re} h (t_1, t_{}, t_3, t_2) \right\}
  \nonumber\\
  u (t, t_1, t_2, t_3) & = & 2 \sum_{\alpha \beta \gamma} \left\{ \tmop{Re} f
  (t - t_{\alpha}) \tmop{Re} f (t_{\beta} - t_{\gamma}) + 2 \tmop{Re} g (t -
  t_{\alpha}) \tmop{Re} f (t_{\beta} - t_{\gamma}) - \tmop{Re} h (t_{\alpha},
  t_{}, t_{\beta}, t_{\gamma}) \right\} \nonumber\\
  &  & - 2 \tmop{Re} h (t, t_1, t_2, t_3) \nonumber\\
  v (t, t_1, t_2, t_3) & = & 2 \sum_{\alpha \beta \gamma} \left\{ f
  (t_{\alpha} - t_{}) f (t_{\gamma} - t_{\beta}) - h (t_{\alpha}, t_{},
  t_{\gamma}, t_{\beta}) \right\}, \nonumber
\end{eqnarray}
where the following summation convention has been used
\begin{eqnarray}
  \sum_{\alpha \beta \gamma} \psi (t_{\alpha}, t_{\beta}, t_{\gamma}) & = &
  \psi (t_1, t_2, t_3) + \psi (t_2, t_1, t_3) + \psi (t_3, t_1, t_2) . 
  \label{eq:sum}
\end{eqnarray}

Including terms up to fourth order one therefore has the expression
Eq.~(\ref{eq:tclnth}) with time dependent coefficients given by the
identifications
\begin{eqnarray}
  \gamma_s (t) & = & -\mathfrak{f}_{} (t)
  -\mathfrak{g} (t) +\mathfrak{p} (t)
  +\mathfrak{r}_{} (t) +\mathfrak{v}_{} (t) 
  \label{eq:gamma}\\
  \gamma_+ (t) & = & 2\mathfrak{f}_R (t)
  +\mathfrak{t} (t) \nonumber\\
  \gamma_+ (t) & = & 2\mathfrak{g}_R (t)
  +\mathfrak{u} (t) \nonumber\\
  \gamma_d (t) & = & \mathfrak{q} (t) +\mathfrak{s}
  (t) + 2\mathfrak{v}_R (t) . \nonumber
\end{eqnarray}

\end{widetext}

\end{document}